\documentclass[doublecol]{epl2}
\usepackage{amsmath}
\usepackage{amssymb}
\usepackage{graphicx}
\usepackage{dcolumn}
\usepackage{bm}
\usepackage{color}
\definecolor{blue}{rgb}{0.3,0.3,0.9}
\usepackage[colorlinks=true,linkcolor=red,anchorcolor=red,citecolor=blue,urlcolor=blue]{hyperref}
\usepackage{graphicx}
\usepackage{color}
\usepackage{hyperref,url}
\usepackage{epstopdf}

\newcommand{\scrh}{{\mathcal H}}

\title{Bosonic Haldane insulator in the presence of local disorder: A quantum Monte Carlo study}

\author{Jian-Ping Lv\inst{1,2}\footnote{phys.lv@gmail.com} \and Jian-Sheng Wang\inst{1}\footnote{phywjs@nus.edu.sg}}
\shortauthor{J.-P. Lv \& J.-S. Wang}
\institute{
  \inst{1} Department of Physics, National University of Singapore, Singapore 117551, Republic of Singapore\\
  \inst{2} Department of Physics, Anhui Normal University, Wuhu 241000, P. R. China
}

\date{March 20, 2018}


\abstract{
The Haldane phase (HP) is a paradigmatic example of symmetry protected topological phase. We explore how the bosonic HP behaves in the presence of local disorder, employing quantum Monte Carlo simulations of an extended Bose-Hubbard model subject to uncorrelated, quenched disorders. We find that the HP is robust against a weak disorder and the non-local string order of HP exhibits a reentrant behavior. Besides, a direct transition between the HP and superfluid phase is uncovered. A significant part of the ground-state phase diagram is established for the model, unveiling the location of HP surrounded by Bose glass, charge density wave and superfluid phases. We also mention a possible experimental scheme with optical lattice emulator to realize the present findings.}

\begin{document}

\maketitle

{\it Introduction}\,---\,The subtle interplay between interaction and disorder leads to a variety of exotic mechanisms such as Anderson localization~\cite{PhysRev.109.1492} and order by disorder~\cite{villain1980order,PhysRevLett.62.2056,henley1987ordering}, challenging prior comprehension towards collective behaviors of many-body systems. The topological phase of interacting particles
is currently at the very heart of condensed matter research~\cite{hasan2010colloquium,qi2011topological} and provides a novel platform to examine general laws of disorder effects.

The celebrated Haldane phase, initially conjectured for integer spin chains~\cite{PhysRevLett.50.1153}, is a prominent example of symmetry protected topological (SPT) phase~\cite{PhysRevB.80.155131,PhysRevB.81.064439,pollmann2012symmetry}.
Bosonic Haldane insulator (BHI) shares some similar characteristics with the Haldane phase of the spin-$1$ Heisenberg antiferromagnetic chain and  the AKLT Hamiltonian~\cite{Affleck1987}, and was first found in an extended Bose-Hubbard (EBH) model with long-range repulsive interactions~\cite{dalla2006hidden,berg2008rise}. The minimal model for the BHI is probably the one-dimensional EBH model of both contact ($U$) and nearest-neighbour ($V$) repulsions, hosting a hidden BHI that indicates the compromise of competitions between Coulomb interactions and kinetic energy~\cite{rossini2012phase,batrouni2013competing,batrouni2014competing}. As a rare example of Haldane phase hosted by a well-known model of interacting bosons, the BHI deserves extensive further research, not only because of the fundamental importance itself, but also due to the possibility of its realization in optical lattices~\cite{landig2015quantum}. A hallmark of SPT phase is the finite bulk gap together with the degenerate or gapless edge excitation protected by a symmetry. Recently, a lowest entanglement level of fourfold degeneracy protected by lattice inversion symmetry has been confirmed in BHI~\cite{ejima2014spectral,PhysRevB.91.045121}. However, to our best knowledge, the disorder effects on BHI, which should be a crucial aspect of a SPT phase, have not been well explored.
\begin{figure}
\includegraphics[angle=0,width=8cm,height=6cm]{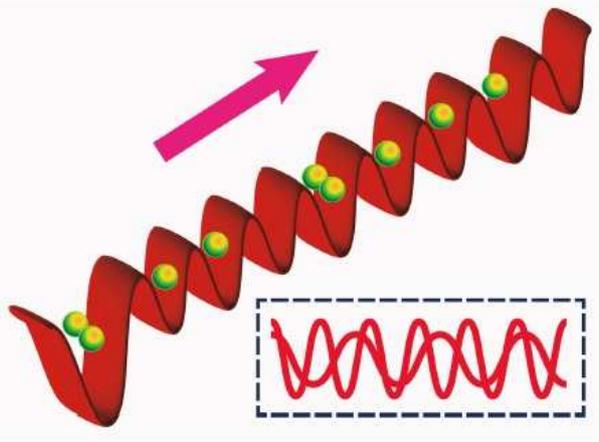}
\caption{(Color online) A schematic illustration of BHI phase on a one-dimensional optical lattice with random potentials. The inset depicts how to produce a disordered optical lattice using optical sublattices of incommensurate wavelengths.}
\label{ol}
\end{figure}

{\it Model}\,---\,Aiming to answer the question how quenched disorders act on a BHI, we investigate the one-dimensional EBH Hamiltonian subject to quenched disorders in the form of random local potentials,
\begin{eqnarray}
\hat{\scrh}&=&-t \sum\limits_{\langle ij \rangle } (\hat{a}_{i}^\dagger \hat{a}_{j} + \hat{a}_{j}^\dagger \hat{a}_{i}) + \frac{U}{2} \sum\limits_{i}  \hat{n}_i(\hat{n}_i-1) \nonumber  \\
&&+ V \sum\limits_{\langle ij \rangle} \hat{n}_i \hat{n}_j+ \sum\limits_{i} \epsilon_i  \hat{n}_i,
\label{BH}
\end{eqnarray}
where $\hat{a}_i^\dagger$ ($\hat{a}_i$) creates (annihilates) a scalar boson at site $i=1,...,L$; $\hat{n}_i=\hat{a}_i^\dagger \hat{a}_i$; $t$ denotes the hopping amplitude; $\langle ij \rangle$ represents nearest-neighbour pairs, and periodic boundary condition is used. 
Uncorrelated disorder is incorporated by randomly distributing $\epsilon_i$ within [$-\mu-\Delta$,  $-\mu+\Delta$], with  $\Delta$ a finite bound representing disorder strength and $\mu$ the chemical potential only relevant in grand canonical simulations. In what follows we shall set $U \equiv 1$ and $k_B \equiv 1$ for convenience.

The BHI appears at unitary filling where the key physics is captured by local fillings
$0$, $1$ and $2$, which are analogous to the three spin variables of a spin-$1$ model,
with $\hat{n}_i \rightarrow \hat{S}^z_i +1$. Figure~\ref{ol} illustrates the existence of BHI in an optical lattice with random local potentials. The absence of BHI at higher integer fillings has been examined recently~\cite{batrouni2013competing,batrouni2014competing}.
We therefore focus on the averaged particle density $\rho \equiv \frac{1}{L} [\langle \sum_{i}\hat{n}_i \rangle]=1$,  with $\langle\cdots\rangle$ denoting thermal average for a given disorder realization and $[\cdots]$ being over realizations of a fixed disorder strength. Most simulations, if not particularly indicated, are performed in the canonical ensemble. Additional grand-canonical simulations are mainly for an unbiased measurement of compressibility via particle number fluctuations.

{\it Method}\,---\,The task to simulate a system in the presence of quenched disorder within the framework
of Monte Carlo method, requiring averagings over both thermal and quenched disorders, is computationally time-consuming. We employ an efficient worm quantum Monte Carlo algorithm~\cite{prokof1998exact,prokof1998worm}, which is unbiased in the sense that it is based on a state space of {\it continuous} imaginary time world-line configurations. This original state space is enlarged by a pair of defects whose walks lead to configuration updates, with samplings being taken in the original space. Local transition probabilities are allocated according to the standard Metropolis-Hastings scheme. Approaching to the computation limit of state-of-the-art quantum Monte Carlo simulations, the maximum chain length and maximum inverse temperature (lowest temperature) we consider are $L_{\rm max}=512$ and $\beta_{\rm max}= 1/T_{\rm min}=320$, respectively.

\begin{figure}
\includegraphics[angle=0,width=8cm,height=7.2cm]{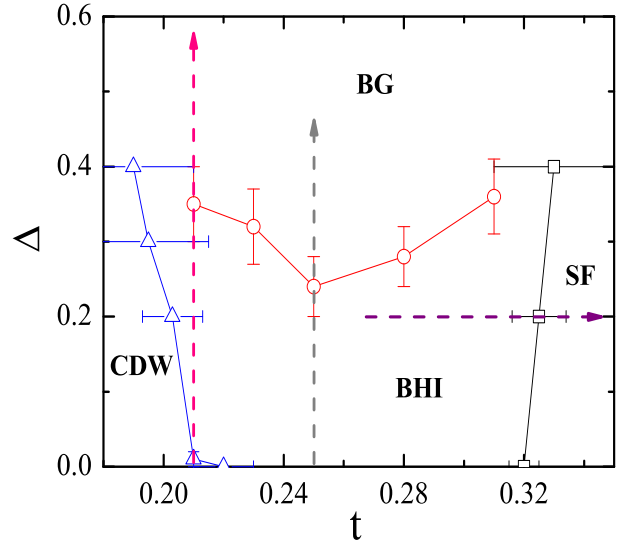}
\caption{(Color online) The $t-\Delta$ ground-state phase diagram of model~(\ref{BH}) at the unitary filling for a moderate interaction ratio $V=3/4$. It contains a BHI phase surrounded by the superfluid (SF), CDW and Bose glass (BG) phases. The symbols are determined by quantum Monte Carlo simulations and the solid lines serve as eyes' guide. The gray, pink, purple dashed paths will be described by
Figs.~\ref{t0.25}, \ref{t0.21} and \ref{dt}, respectively.}
\label{pd}
\end{figure}

The superfluid density $\rho_{\rm s}$ is sampled through winding number ($W$) fluctuations over world-line configurations with~\cite{PollockCeperley}
\begin{equation}\label{DR}
\rho_{\rm s}=  \frac{L [\langle W^2 \rangle]}{2\beta t}.
\end{equation}
Simultaneously, the charge density wave (CDW) order and the non-local string order are characterized through diagonal correlation functions
\begin{equation}\label{DOC}
\mathrm{O_c}(r)=(-1)^{|i-j|}[\langle \delta \hat{n}_i \delta \hat{n}_j \rangle]  \\
\end{equation}
and
\begin{equation}\label{DOS}
\mathrm{O_s}(r)=[\langle \delta \hat{n}_i e^{i \theta \Sigma^{j-1}_{k=i} \delta \hat{n}_k }\delta \hat{n}_j \rangle],  \\
\end{equation}
respectively. Here $r$ means the intersite distance between site $i$ and $j$ along the periodic chain, and $\delta \hat{n}_i=\hat{n}_i-1$ is the local particle number difference with respect to the unitary filling. As confirmed in Ref.~\cite{qin2003nonlocal}, $\theta=\pi$ is a well-defined topological angle for string order~\cite{den1989preroughening}. The measurement of correlations is taken as a function of intersite distance, with special attention being paid to the longest distance along a finite chain\,---\,for periodic chains, we focus on $r=L/2$.
To distinguish between different quantum glass phases, we sample the compressibility given by
\begin{equation}\label{DC}
\chi = \frac{\beta}{L}[\langle (\sum_{i}\hat{n}_i)^2 \rangle- \langle \sum_{i}\hat{n}_i \rangle^2],
\end{equation}
using additional simulations in the grand-canonical ensemble.

\begin{table}\caption{The main criteria we use to distinguish among the SF, CDW, BHI, BG and Mott glass (MG) phases. The relevant quantities are defined by Eqs.~(\ref{DR}), (\ref{DOC}), (\ref{DOS}) and (\ref{DC}).}\label{phases}
\centering
\begin{tabular}{ccccc}
  \hline
  \hline
        & $\rho_{\rm s}$ &$\mathrm{O_c}(L/2)$ &$\mathrm{O_s}(L/2)$   & $\chi$ \\
      \hline

     SF & $\neq 0$  & $=0$  & $=0$  & $\neq 0$ \\
     CDW & $=0$ & $\neq 0$  & $\neq 0$   & $=0$ \\
     BHI & $=0$ & $=0$  & $\neq 0$   & $=0$\\
     BG & $=0$ & $=0$  & $= 0$  & $\neq 0$ \\
     MG & $=0$ & $=0$  & $= 0$  & $=0$ \\
  \hline
  \hline
\end{tabular}
\end{table}

{\it Ground-state phase diagram}\,---\,The quantum phases on a characteristic ground-state phase diagram (Fig.~\ref{pd}) are identified according to the criteria listed in Table~\ref{phases}, by taking the limits $L \rightarrow \infty $ and $\beta \rightarrow \infty$ of the size- and temperature-dependent Monte Carlo data. The phase diagram contains SF, BHI, CDW and BG phases, partially reflecting the stability of BHI phase against quenched disorders. In a visible domain of hopping amplitude, the BHI phase remains stable at a moderate disorder strength $\Delta \sim 0.3$. Doping a CDW phase, one may find that the crystalline order is less stable than the non-local string order and a CDW-BHI transition happens. Moreover, a reentrant behavior of the non-local string order is observed in the BHI. As a result, the BHI regime in the presence of weak disorder is broader than that of clean case. Concrete evidences from quantum Monte Carlo simulations will be detailed below.

\begin{figure}
\includegraphics[angle=0,width=8cm,height=12cm]{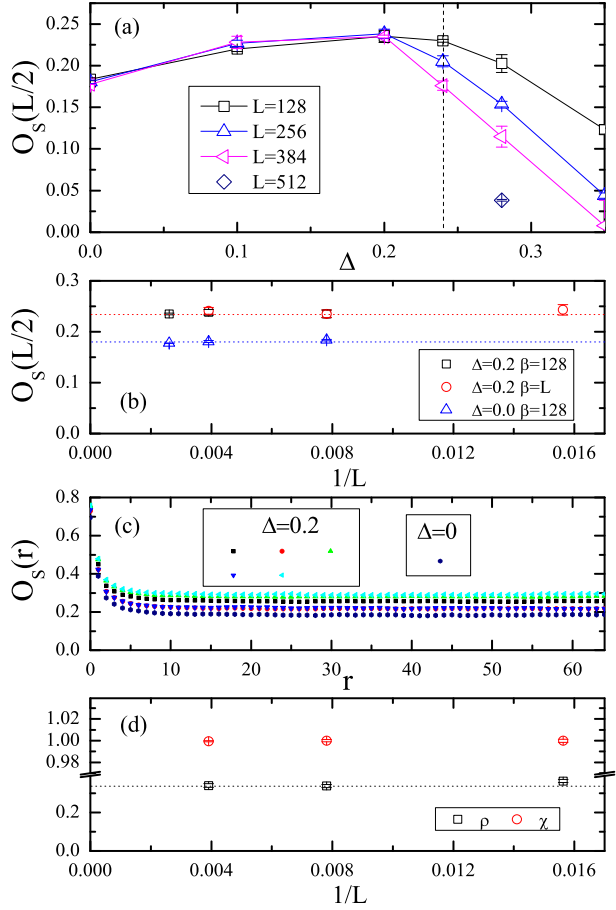}
\caption{(Color online) (a) Diagonal correlations $\mathrm{O_s}(L/2)$ and $\mathrm{O_c}(L/2)$ versus $\Delta$ for the parameter set ($t=1/4$, $V=3/4$, $\beta=128$) which corresponds to a BHI phase at $\Delta=0$. Here and in the following plots, error bars are evaluated from sample-to-sample fluctuations and statistical. The dashed line indicates a BHI-BG transition. (b) Finite-size scalings of $\mathrm{O_s}(L/2)$ at $\Delta=0$ and $0.2$. (c) String order correlation function $\mathrm{O_s}(r)$ versus intersite distance with the parameter set ($t=1/4$, $V=3/4$, $\beta=L=128$) for different disorder realizations of $\Delta=0.2$. The same correlation function for the clean case $\Delta=0$ is also plotted for comparison. (d) Finite-size scaling on $\chi$ of a BG phase with the parameter set ($t=1/4$, $V=3/4$, $\Delta=0.35$). The data are obtained by grand-canonical simulations with $\mu$ being optimally adjusted such that $\rho \approx 1$. The actual particle density $\rho$ is shown as well. In this subplot, the inverse temperatures are set as $\beta=L$.}
\label{t0.25}
\end{figure}

{\it Quantum Monte Carlo results}\,---\,We begin with a straightforward examination on the stability of BHI in the presence of quenched disorder. The parameter set we take as an example is ($V=3/4$, $t=1/4$) which, at the clean case $\Delta=0$, corresponds to a deep BHI phase~\cite{batrouni2013competing,lv2015exotic}. Shown in Fig.~\ref{t0.25}(a), as $\Delta$ increases, the string order parameter $\mathrm{O_s}(L/2)$ increases first and then, at a transition point $\Delta_c=0.24(4)$, drops drastically to zero. Not surprisingly, the correlation $\mathrm{O_c}(L/2)$ keeps vanishing for every considered $\Delta$ in the thermodynamic limit, despite minor corrections that quickly shrink as $L$ increases (not shown). Figure~\ref{t0.25}(b) shows finite-size scalings of $\mathrm{O_s}(L/2)$ for $\Delta=0$ and $0.2$. Firstly, the discrepancies of extrapolations inferred from different scalings are negligible, and the extrapolations to $L = \infty$ are quickly saturated. Further, as the disorder strength increases from $\Delta=0$ to $0.2$, we confirm an obvious enhancement of $\mathrm{O_s}(L/2)$. As a closer look at the enhanced string order, we show in Fig.~\ref{t0.25}(c) string order correlation functions $\mathrm{O_s}(r)$ for different $\Delta$. For $\Delta=0$ and for every disordered realization of $\Delta=0.2$, the string order correlation remains finite as $r$ increases. Moreover, as $r$ varies, $\mathrm{O_s}(r)$ is almost monotonic in the sense that there is no obvious modulation as we often find for $\mathrm{O_c}(r)$ in a CDW phase. Clear evidence is also obtained to support that, comparing to the clean case ($\Delta=0$), most realizations for $\Delta=0.2$ exhibit an enhancement of string order correlation in the long-range limit. We therefore conclude that the BHI phase is stable under weak quenched disorder, and more intriguingly, the string order in a BHI can be enhanced by the disorder. The latter is known as reentrant behavior.

In the strongly disordered case, $\Delta>\Delta_c$, we find a quantum glass phase that is insulating and with neither long-range nor quasi-long-range ordering, namely, $\mathrm{O_s}(L/2)=0$, $\mathrm{O_c}(L/2)=0$ and $\rho_{\rm s}=0$. The glass phase deserves a further analysis. There are in general two scenarios of glass phases of bosons, BG phase~\cite{giamarchi1987localization,giamarchi1988anderson,fisher1989boson} and MG phase~\cite{PhysRevB.64.245119,PhysRevLett.93.150402,PhysRevLett.92.015703}, where a crucial difference is the existence of a finite compressibility.
For the glass phase at $t=1/4$, $V=3/4$ and $\Delta=0.35$, we perform grand canonical simulations, adjusting the chemical potential to achieve a good approximation of $\rho=1$. Typically the optimal particle density can be as precise as $\rho \approx 1.000$.
Aforementioned physical quantities of interest are simultaneously sampled. Figure~\ref{t0.25}(d) demonstrates that $\chi$ extrapolates fast to a finite value as $L$ increases (the limit to $T=0$ is simultaneously taken). As we have confirmed all the characteristics listed in Table~\ref{phases} for a BG phase, we infer that the disordered phase with $\Delta > \Delta_c$ is a BG phase, albeit it locates near the boundary of the incompressible, gapped BHI phase. It follows that, as the quenched disorder is incorporated, the BHI phase can be destroyed by closing the gap, in accompany with the emergence of a finite compressibility.

 We examine the possible self-averaging properties of BHI and BG phases, by estimating the numbers of disorder realizations required to
 characterize each phase. We take the present precision as a benchmark. For the BHI phase a relatively small number ($\sim 100$) of realizations is sufficient to reveal key features such as the string order enhancement, while a quantitative evaluation of BG phase requires a considerably larger number\,---\,even for a moderate disorder strength $\Delta \lesssim 0.5$ we require realizations of a typical number of $\sim 1000$. Therefore, comparing to the BG phase, the BHI phase exhibits a more sizeable self-averaging effect.

\begin{figure}
\includegraphics[angle=0,width=8cm,height=7cm]{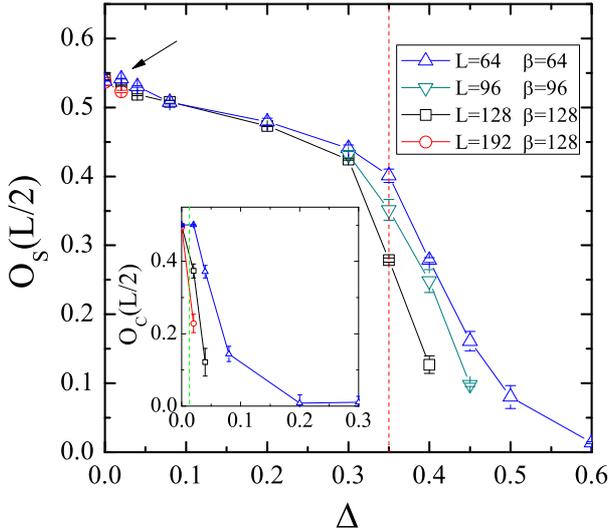}
\caption{(Color online) String order correlation $\mathrm{O_s}(L/2)$ versus $\Delta$ for the parameter set ($t=0.21$, $V=3/4$). The inset is for $\mathrm{O_c}(L/2)$ and focuses on small $\Delta$.  The red (green) dashed line indicates a BHI-BG (CDW-BHI) transition, and the arrow points to a kink corresponding to the CDW-BHI transition.}
\label{t0.21}
\end{figure}

To compare the stabilities of typical diagonal orders\,---\,the CDW order and the non-local string order\,---\,in the vicinity of the CDW-BHI phase boundary, we compute the quantities $\mathrm{O_c}(L/2)$ and $\mathrm{O_s}(L/2)$ for a wide range of $\Delta$. The example shown in Fig.~\ref{t0.21} is for the parameter set ($V=3/4$, $t=0.21$) whose clean limit $\Delta=0$ corresponds to a CDW phase where $\mathrm{O_c}(L/2)$ and $\mathrm{O_s}(L/2)$ are both robust. It is clear that the CDW order $\mathrm{O_c}(L/2)$ is rather unstable and dissolves at a disorder strength as small as $\Delta_{c,1}=0.01(1)$. By contrast, the string order $\mathrm{O_s}(L/2)$ persists up to a considerably strong disorder strength. The transition point from CDW to BHI is thus estimated as $\Delta_{c,1}$, whereas a BHI-BG transition happens at $\Delta_{c,2}=0.35(5)$. In the present situation, the non-local string order is more stable than the crystalline CDW order. These results correspond to the negative slope of CDW-BHI boundary in the ground state phase diagram (Fig.~\ref{pd}).

\begin{figure}
\includegraphics[angle=0,width=8cm,height=8cm]{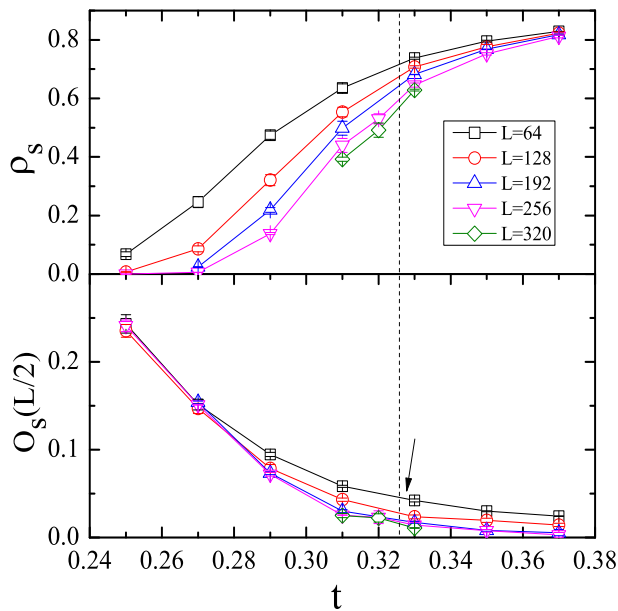}
\caption{(Color online) SF density $\rho_{\rm s}$ and string order correlation $\mathrm{O_s}(L/2)$ versus $t$ for different $L$ with $V=3/4$, $\Delta=0.2$ and $\beta=L$. The dashed line indicates a BHI-SF transition, and the arrow points to a kink at the transition point.}
\label{dt}
\end{figure}

Is there a direct BHI-SF transition at a finite $\Delta$? Substantial research efforts have been devoted to a similar question\,---\,the existence of a direct transition between Mott insulator and SF in the disordered Bose-Hubbard model, where the current consensus is a negative answer~\cite{weichman2008dirty}. While both BHI and Mott insulator phases are gapped, the former is a SPT phase and the latter is topologically trivial. Intriguingly, because of the stability of BHI, it is likely that a direct BHI to SF transition survives, without a sandwiching BG region. We show in Fig.~\ref{dt} for a moderate disorder strength $\Delta=0.2$ that the transition point inferred from $\rho_{\rm s}$ for the insulator-SF transition is $t_c=0.325(9)$ where $\rho_{\rm s}$ exhibits a profound variation. Beyond the range indicated by error bar, the superfluidity density is either robust or vanishing. The precision of estimated transition point is therefore confirmed. Rightly at $t_c$, the string order $\mathrm{O_s}(L/2)$ begins to dissolve and shows a drastic drop that is visible for large systems. Therefore, based on the very precise limit of the present Monte Carlo data, together with an overall perspective on the location of BHI-BG transition line (which does not terminate at the $\Delta=0$ axis), we infer that the insulator-SF transition occurs simultaneously with the fading away of non-local string order, and the BG phase does not show up in between. However, given that the BHI features a broken, hidden $\mathbb{Z}_2 \times \mathbb{Z}_2$ symmetry, we can not readily presume that the BHI-SF transition is of the Berezinskii-Kosterlitz-Thouless type.

{\it Summary and discussions}\,---\,In this work, we have shown that the non-local string order of HP exhibits an unusual reentrant behavior\,---\,it can be enhanced by incorporating quench disorder ($\Delta=0$ $\rightarrow$ $\Delta=0^{+}$) or by increasing disorder strength ($\Delta$ $\uparrow$) within a domain. Note that a similar reentrant behaviour was also observed for the phase stiffness of a quantum XY model driven by thermal fluctuations~\cite{Capriotti2003} rather than by quenched disorders. As the competing interactions are present, the BHI phase emerges when disordering a CDW phase\,---\,the crystalline order is destroyed and the non-local string order persists. We found a BG phase bordering the gapped, topologically nontrivial BHI phase on the ground-state phase diagram. The BHI exhibits a more sizeable self-averaging effect comparing to the BG. A direct quantum phase transition between BHI and SF, without a sandwiching region hosting any type of glass phase, was uncovered in presence of quenched disorder.

Hamiltonian~(\ref{BH}) may be realized in a disordered optical lattice, as sketched in Fig.~\ref{ol}.
The realization of a disordered optical lattice is within state-of-the-art experimental capabilities~\cite{disorderedOL1,disorderedOL2}. To achieve random potentials, one can use speckle light patterns and image them onto the Bose-Einstein condensate, as first demonstrated in Ref.~\cite{boiron1999trapping}. Further experimental success was reflected by speckle potentials with a less-than-1$\mu m$ autocorrelation length~\cite{clement2006experimental}. An alternative approach which may be more controllable is to produce a multi-chromatic optical lattice, formed by optical sublattices of incommensurate wavelengths~\cite{fallani2007ultracold}. On the other hand, recent experiments demonstrated a remarkable progress on tuning interactions at different length scales and determined a variety of quantum phases and phase transitions~\cite{landig2015quantum}. Therefore, the present findings may be experimentally achievable. Last but not least, it is of both experimental and theoretical interest to explore disorder effects on the BHI in the presence of long-range interactions and on the universality class of BHI-related phase transitions predicated by a field theory~\cite{dalla2006hidden,berg2008rise}.

\begin{acknowledgments} {\it Acknowledgements}\,---\,We thank Z. D. Wang for fruitful discussions during the collaboration on related topics, and Xiao Huo for the technical help with drawing Fig.~\ref{ol}. We also acknowledge National Supercomputing Center in Shenzhen (Shenzhen Cloud Computing Center), High-Performance Computing Team at NUS Computer Centre, and NUS Physics Department HPC for help with the numerical simulations. This work is supported by FRC grant R-144-000-343-112, MOE grant R-144-000-349-112, and NSFC (Grant Nos. 11774002 and 11405003).
\end{acknowledgments}

\end{document}